\documentclass{article}
\usepackage{amsmath}
\usepackage{amssymb}
\usepackage{amsthm}
\usepackage{stmaryrd}
\usepackage{algorithm}
\usepackage{algorithmic}
\usepackage[pdftex]{graphicx}
\begin{document}
\title{A Time Lower Bound for Multiple Nucleation on a Surface}
\author{Aaron Sterling\footnote{Laboratory for Nanoscale Self-Assembly, Department of Computer Science, Iowa State University, Ames, IA 50014, USA.  \texttt{sterling@cs.iastate.edu}.  This research was supported in part by National Science Foundation Grants 0652569 and 0728806.}}
\maketitle

\newtheorem{definition}{Definition}
\newtheorem{thm}{Theorem}
\newtheorem{prop}{Proposition}
\newtheorem{cor}{Corollary}
\begin{abstract}
Majumder, Reif and Sahu have presented a stochastic model of reversible, error-permitting, two-dimensional tile self-assembly, and showed that restricted classes of tile assembly systems achieved equilibrium in (expected) polynomial time.  One open question they asked was how much computational power would be added if the model permitted multiple nucleation, \emph{i.e.,} independent groups of tiles growing before attaching to the original seed assembly.  This paper provides a partial answer, by proving that if a tile assembly model uses only local binding rules, then it cannot use multiple nucleation on a surface to solve certain ``simple'' problems in constant time (time independent of the size of the surface).  Moreover, this time bound applies to macroscale robotic systems that assemble in a three-dimensional grid, not just to tile assembly systems on a two-dimensional surface.  The proof technique defines a new model of distributed computing that simulates tile (and robotic) self-assembly.  \emph{Keywords:} self-assembly, multiple nucleation, locally checkable labeling.
\end{abstract}
\section{Introduction}
\subsection{Overview}
Nature is replete with examples of the self-assembly of individual parts into a more complex whole, such as the development from zygote to fetus, or, more simply, the replication of DNA itself.  In his Ph.D. thesis in 1998, Winfree proposed a formal mathematical model to reason algorithmically about processes of self-assembly~\cite{winfree}.  Winfree connected the experimental work of Seeman~\cite{seeman} (who had built ``DNA tiles,'' molecules with unmatched DNA base pairs protruding in four directions, so they could be approximated by squares with different ``glues'' on each side) to a notion of tiling the integer plane developed by Wang in the 1960s~\cite{wang}.  Rothemund, in his own Ph.D. thesis, extended Winfree's original Tile Assembly Model~\cite{rothemund}.

Informally speaking, Winfree effectivized Wang tiling, by requiring a tiling of the plane to start with an individual \emph{seed tile} or a connected, finite \emph{seed assembly}.  Tiles would then accrete one at a time to the seed assembly, growing a \emph{seed supertile}.  A \emph{tile assembly system} is a finite set of \emph{tile types}.  Tile types are characterized by the names of the ``glues'' they carry on each of their four sides, and the binding strength each glue can exert.  We assume that when the tiles interact ``in solution,'' there are infinitely many tiles of each tile type.  Tile assembly proceeds in discrete stages.  At each stage $s$, from all possibilities of tile attachment at all possible locations (as determined by the glues of the tile types and the binding requirements of the system overall), one tile will bind, with tile type and location ``chosen'' nondeterministically from possible legal bonds at that stage.  (Later, we will generalize this so multiple tiles can bind concurrently, at a given stage.)  Winfree proved that his Tile Assembly Model is Turing universal.

The abstract Tile Assembly Model (aTAM) is \emph{error-free} and \emph{irreversible}---tiles always bind correctly, and, once a tile binds, it can never unbind.  Adleman \emph{et al.} were the first to define a notion of time complexity for tile assembly, using a one-dimensional error-permitting, reversible model, where tiles would assemble in a line with some error probability, then be scrambled, and fall back to the line~\cite{adcheng}.  Adleman \emph{et al.} proved bounds on how long it would take such models to achieve equilibrium.  Majumder, Reif and Sahu have recently presented a two-dimensional stochastic model for self-assembly~\cite{reif}, and have shown that some tiling problems in their model correspond to \emph{rapidly mixing Markov chains}---Markov chains that reach stationary distribution in time polynomial in the state space of legally reachable assemblies.

While the aTAM is nondeterministic, real-world chemical reactions are probabilistic, and discrete molecular interactions are often modeled stochastically.  We will define a class of stochastic self-assembly models that contains the model of Majumder \emph{et al.}, and prove a lower bound about any model in that class.

The tile assembly systems analyzed in~\cite{reif} had the property that their equilibrium assemblies were identical (allowing for small error) with their \emph{terminal} or \emph{complete assemblies}, \emph{i.e.}, assemblies that cannot legally evolve further, given the rules of the system.  This identity does not, however, hold in general.  In a closed chemical system, where equilibrium may be achieved, it is possible that the system at equilibrium might consist almost entirely of large, undesirable assemblies that do not perform the desired computation.  In these cases, correct assembly occurs when the system is out of equilibrium, and can be maintained because there is a large kinetic energy barrier to forming undesired structures.  Therefore, when we discuss the ``solution to a problem'' in this paper, we identify that with the notion of a complete assembly.

We will prove a time complexity lower bound on the solution of a graph coloring problem for a class of self-assembly models, including, but not limited to, a generalization of the model of~\cite{reif}.  The tile assembly model in~\cite{reif}, like the aTAM, allows only for a single seed assembly, and one of the open problems in~\cite{reif} was how the model might change if it allowed multiple nucleation, \emph{i.e.}, if multiple supertiles could build independently before attaching to a growing seed supertile.  The main result of this paper provides a time complexity lower bound for a class of tile assembly models that permit multiple nucleation on a 2D surface or a 3D grid: there is no way for those models to use multiple nucleation to achieve a speedup to tiling a surface in constant time (time independent of the size of the surface) in order to solve a graph coloring problem, even though that graph coloring problem requires only seven tile types to solve in the aTAM.  This result holds for tile assembly models that are reversible, irreversible, error-permitting or error-free.  In fact, a speedup to constant time is impossible, even if we relax the model to allow that, at each step $s$, there is a positive probability for every available location that a tile will bind there (instead of requiring that exactly one tile bind per stage).

To our knowledge, the method of proof in this paper is novel: given a tile assembly model and a tile assembly system $\mathcal{T}$ in that model, we construct a distributed network of processors that can simulate the behavior of $\mathcal{T}$ as it assembles on a surface.  Our result then follows from the theorem by Naor and Stockmeyer that locally checkable labeling (LCL) problems have no local solution in constant time~\cite{ns}.  This is true for both deterministic and randomized algorithms, so no constant-time tile assembly system exists that solves an LCL problem with a positive probability of success.  We consider one LCL problem in specific, the weak $c$-coloring problem, and demonstrate a tile set of only seven tile types that solves the weak $c$-coloring problem in the abstract Tile Assembly Model, even though that same problem is impossible to solve in constant time by multiple nucleation on a surface, for a broad class of self-assembly models.  Intuitively, this demonstrates that even a problem that can be solved in polynomial time by using a few local rules when starting from a single point, cannot necessarily be solved in constant time when starting from multiple points, regardless of the rule set used.  (The abstract Tile Assembly Model can weakly $c$-color an $n \times n$ surface in $n^2$ steps, yet none of the multiple nucleation models we consider can solve the weak $c$-coloring problem in constant-many steps.)

The results of Naor and Stockmeyer we apply are more powerful than needed to obtain the time complexity lower bound for a system in which the self-assembling agents are as simple as DNA tiles.  Our lower bound actually demonstrates that constant-time speedup to solve LCL problems is impossible via multiple nucleation, even for self-assembling modular robots capable of forming physical bonds in a three-dimensional grid, and, in addition, of sending messages to their neighbors once they have bonded, and potentially deciding to break bonds they previously formed.
\subsection{Background}
In the abstract Tile Assembly Model, one tile is added per stage, so the primary complexity measure is not one of time, but of how much information a tile set needs in order to solve a particular problem.  Several researchers~\cite{adcheng}~\cite{complexities}~\cite{openproblems}~\cite{programsize}~\cite{solwin} have investigated the \emph{tile complexity} (the minimum number of distinct tile types required for assembly) of finite shapes, and sets of ``scale-equivalent'' shapes (essentially a $\mathbb{Z}\times\mathbb{Z}$ analogue of the Euclidean notion of similar figures).  For example, it is now known that the number of tile types required to assemble a square of size $n \times n$ (for $n$ any natural number) is $\Omega(\log n / \log \log n)$~\cite{programsize}. Or, if $T$ is the set of all discrete equilateral triangles, the asymptotically optimal relationship between triangle size and number of tiles required to assemble that triangle, is closely related to the Kolmogorov Complexity of a program that outputs the triangle as a list of coordinates~\cite{solwin}.

Despite these advances in understanding of the complexity of assembling finite, bounded shapes, the self-assembly of infinite structures is not as well understood.  In particular, there are few lower bounds or impossibility results on what infinite structures can be self-assembled in the Tile Assembly Model.  The first such impossibility result appeared in~\cite{ssadst}, when Lathrop, Lutz and Summers showed that no finite tile set can assemble the discrete Sierpinski Triangle by placing a tile only on the coordinates of the shape itself.  (By contrast, Winfree had shown that just seven tile types are required to tile the first quadrant of the integer plane with tiles of one color on the coordinates of the discrete Sierpinski Triangle, and tiles of another color on the coordinates of the complement~\cite{winfree}.)  Recently, Patitz and Summers have extended this initial impossibility result to other discrete fractals~\cite{patsum}, and Lathrop \emph{et al.}~\cite{ccsa} have demonstrated sets in $\mathbb{Z} \times \mathbb{Z}$ that are Turing decidable but cannot be self-assembled in Winfree's sense.

To date, there has been no work comparing the strengths of different tile assembly models with respect to infinite (nor to finite but arbitrarily large) structures.  Since self-assembly is a process in which each point has only local knowledge, it is natural to consider whether the techniques of distributed computing might be useful for comparing models of self-assembly and proving impossibility results about them.  This paper is an initial attempt in that direction.

Aggarwal \emph{et al.} in~\cite{complexities} proposed a generalization of the standard Tile Assembly Model, which they called the $q$-Tile Assembly Model.  This model permitted multiple nucleation: tiles did not need to bind immediately to the seed supertile.  Instead, they could form independent supertiles of size up to some constant $q$ before then attaching to the seed supertile.  While the main question considered in~\cite{complexities} was \emph{tile} complexity, we can also ask whether multiple nucleation would allow an improvement in \emph{time} complexity.  Intuitively, Does starting from multiple points allow us to build things strictly faster than starting from a single point?

As mentioned above, Majumder, Reif and Sahu recently presented a stochastic, error-permitting tile assembly model, and calculated the rate of convergence to equilibrium for several tile assembly systems~\cite{reif}.  The model in~\cite{reif} permitted only a single seed assembly, and addition of one tile to the seed supertile at each stage.  Majumder, Reif and Sahu left as an open question how the model might be extended to permit the presence and binding of multiple supertiles.

Therefore, we can rephrase the ``intuitive'' question above as follows: Can we tile a surface of size $n \times n$ in a constant number of stages, by randomly selecting nucleation points on the surface, building supertiles of size $q$ or smaller from those points in $\leq q$ stages, and then allowing $\leq r$ additional stages for tiles to fall off and be replaced if the edges of the supertiles contain tiles that bind incorrectly?  (The assembly achieves equilibrium in constant time because $q$ and $r$ do not depend on $n$.)  The partial answer obtained in this paper is that locally checkable labeling problems cannot be solved in constant time, if we limit ourselves to self-assembly on a surface.

Limiting ourselves to self-assembly on a surface is significant, because we are requiring that agents adhere to a substrate and then never move again, unless they dissociate completely from the larger assembly.  When assemblies multiply nucleate in solution, however, they form disjoint supertiles that can float independently until potentially becoming aligned, with some probability.  A self-assembly model that made this rigorous might be strictly stronger than the self-assembly models we consider in this paper, as it is not clear how to simulate floating supertiles within our distributed computing models without introducing slowdown, as processors simulating locations of the surface would have to ``pass along'' information from one processor to the next, to simulate elements of the moving supertile.  We leave the possibility of simulating floating supertiles to future work.

Another limitation to our results is that our proof technique applies only to self-assembly models whose binding rules are completely local.  One could imagine models in which supertiles combine (or separate) based on simultaneous interactions at several locations, instead of the models we consider in this paper, in which the system's behavior at each location depends only on the properties of that location's immediate neighbors.  The self-assembly literature, to our knowledge, contains little regarding self-assembly models with nonlocal binding rules, and this could be a fruitful area to investigate.

Klavins and co-authors have modeled self-assembly phenomena---and programmed self-assembling modular robots---using graph grammars~\cite{klavins:directed}~\cite{klavinsghrist}.  Klavins in~\cite{klavins:psa} informally compares the limitations of the ``distributed algorithms'' of graph grammars (used to program self-assembling robots) to impossibility results in distributed computing.  Recently, we have shown connections between self-assembly and the wait-free consensus hierarchy~\cite{datsa}, and we have embedded the ``graph assembly systems'' of Klavins into a known graph grammar characterization of distributed systems~\cite{sasds}.  The present paper, to the best of our knowledge, is the first to construct a formal reduction from self-assembly models to models of distributed computing.

Section 2 of this paper describes the abstract Tile Assembly Model, and then considers generalizations of the standard model that permit multiple nucleation. Section 3 reviews the distributed computing results of Naor and Stockmeyer needed to prove the impossibility result.  In Section 4 we present our simulation technique and lower bound results.  Section 5 concludes the paper and suggests directions for future research.

\section{Description of Self-Assembly Models}
\subsection{The abstract Tile Assembly Model}
Winfree's objective in defining the Tile Assembly Model was to provide a useful mathematical abstraction of DNA tiles combining in solution~\cite{winfree}.  Rothemund~\cite{rothemund}, and Rothemund and Winfree~\cite{programsize}, extended the original definition of the model.  For a comprehensive introduction to tile assembly, we refer the reader to~\cite{rothemund}.  In our presentation here, we follow~\cite{ssadst}, which gives equal status to finite and infinite tile assemblies.

Intuitively, a tile of type $t$ is a unit square that can be placed with its center on a point in the integer lattice.  A tile has a unique orientation; it can be translated, but not rotated.  We identify the side of a tile with the direction (or unit vector) one must travel from the center to cross that side.  The literature often refers to west, north, east and south sides, starting at the leftmost side and proceeding clockwise.  Each side of a tile is covered with a ``glue'' that has a \emph{color} and a \emph{strength}.  Figure 1 shows how a tile is represented graphically.

\begin{figure}
\centering
\includegraphics[height=3in]{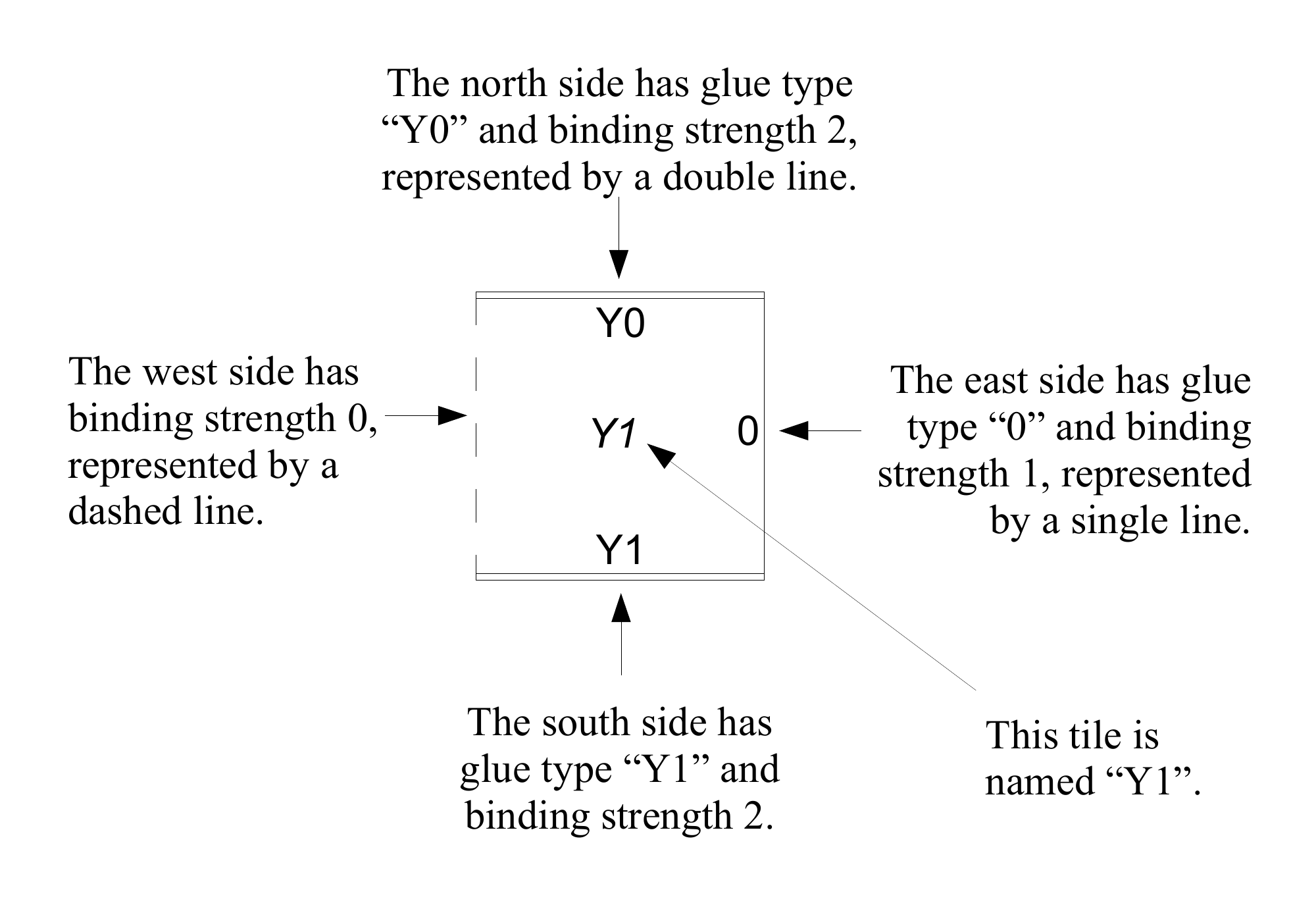}
\caption{An example tile with explanation.} \label{figure:example}
\end{figure}

If tiles of types $t$ and $t'$ are placed adjacent to each other, then they will \emph{bind} with the strength shared by both adjacent sides if the glues on those sides are the same.  Note that this definition of binding implies that if the glues of the adjacent sides do not have the same color or strength, then their binding strength is 0.  Later, we will permit pairs of glues to have negative binding strength, to model error occurrence and correction.

One parameter in a tile assembly model is the minimum binding strength required for tiles to bind ``stably.''  This parameter is usually termed \emph{temperature} and denoted by $\tau$, where $\tau \in \mathbb{N}$.

As we consider only two-dimensional tile assemblies, we limit ourselves to working in $\mathbb{Z}^2=\mathbb{Z} \times \mathbb{Z}$. $U_2$ is the set of all unit vectors in $\mathbb{Z}^2$.

A \emph{binding function} on an (undirected) graph $G=(V,E)$ is a function $\beta : E \longrightarrow \mathbb{N}$.  If $\beta$ is a binding function on a graph $G=(V,E)$ and $C=(C_0,C_1)$ is a cut of $G$, then the \emph{binding strength} of $\beta$ on $C$ is
\begin{displaymath}
\beta_C=\{\beta(e) \mid e \in E, e \in C_0,\textrm{ and } e \in C_1\} \enspace .
\end{displaymath}
The \emph{binding strength} of $\beta$ on $G$ is then $\beta(G)=\min\{\beta_C \mid C \textrm{ is a cut of } G\}$.  Intuitively, the binding function captures the strength with which any two neighbors are bound together, and the binding strength of the graph is the minimum strength of bonds that would have to be severed in order to separate the graph into two pieces.

A \emph{binding graph} is an ordered triple $G=(V,E,\beta)$ where $(V,E)$ is a graph and $\beta$ is a binding function on $(V,E)$.  If $\tau \in \mathbb{N}$, a binding graph $G=(V,E,\beta)$ is \emph{$\tau$-stable} if $\beta((V,E)) \geq \tau$.

Recall that a \emph{grid graph} is a graph $G=(V,E)$ where $V \subseteq \mathbb{Z} \times \mathbb{Z}$ and every edge $\{\overrightarrow{m},\overrightarrow{n}\} \in E$ has the property that $\overrightarrow{m}-\overrightarrow{n} \in U_2$. We write $[V]^2$ for the set $\left\{ \{v_1,v_2\} \mid v_1 \in V \textrm{ and } v_2 \in V \right\}$, \emph{i.e.}, the two-element subsets of $V$.
\begin{definition}
A \emph{tile type} over a (finite) alphabet $\Sigma$ is a function $t : U_2 \longrightarrow \Sigma^* \times \mathbb{N}$. We write $t = (\mathrm{col}_t,\mathrm{str}_t)$, where $\mathrm{col}_t : U_2 \longrightarrow \Sigma^*$, and $\mathrm{str}_t : U_2 \longrightarrow \mathbb{N}$ are defined by $t(\overrightarrow{u}) = (\mathrm{col}_t(\overrightarrow{u}), \mathrm{str}_t(\overrightarrow{u}))$ for all $\overrightarrow{u} \in U_2$.
\end{definition}
\begin{definition}
If $T$ is a set of tile types, a \emph{$T$-configuration} is a partial function $\alpha: \mathbb{Z}^2 \dashrightarrow T$.
\end{definition}
\begin{definition}
The \emph{binding graph} of a $T$-configuration $\alpha:\mathbb{Z}^2 \dashrightarrow T$ is the binding graph $G_{\alpha}=(V,E,\beta)$, where $(V,E)$ is the grid graph given by
\begin{description}
\item $V=\mathrm{dom}(\alpha),$
\item $E=\big\{ \{\overrightarrow{m},\overrightarrow{n}\} \in [V]^2 \mid \overrightarrow{m} - \overrightarrow{n} \in U_2, \mathrm{col}_{\alpha(\overrightarrow{m})}(\overrightarrow{n}-\overrightarrow{m}) = \mathrm{col}_{\alpha(\overrightarrow{n})}(\overrightarrow{m} - \overrightarrow{n}),$ and $\mathrm{str}_{\alpha(\overrightarrow{m})}(\overrightarrow{n}-\overrightarrow{m})>0 \big\},$
\end{description}
and the binding function $\beta: E \longrightarrow \mathbb{Z}^+$ is given by $\beta(\{\overrightarrow{m},\overrightarrow{n}\}) = \mathrm{str}_{\alpha(\overrightarrow{m})}(\overrightarrow{n}-\overrightarrow{m})$ for all $\{\overrightarrow{m},\overrightarrow{n}\} \in E$.
\end{definition}
\begin{definition}
For $T$ a set of tile types, a $T$-configuration $\alpha$ is \emph{stable} if its binding graph $G_{\alpha}$ is $\tau$-stable.  A \emph{$\tau$-$T$-assembly} is a $T$-configuration that is $\tau$-stable.  We write $\mathcal{A}^{\tau}_T$ for the set of all $\tau$-$T$-assemblies.
\end{definition}
\begin{definition}
Let $\alpha$ and $\alpha '$ be $T$-configurations.
\begin{enumerate}
\item $\alpha$ is a \emph{subconfiguration} of $\alpha'$, and we write $\alpha \sqsubseteq \alpha'$, if $\mathrm{dom}(\alpha) \subseteq \mathrm{dom}(\alpha')$ and, for all $\overrightarrow{m} \in \mathrm{dom}(\alpha)$, $\alpha(\overrightarrow{m}) = \alpha'(\overrightarrow{m})$.
\item $\alpha'$ is a \emph{single-tile extension} of $\alpha$ if $\alpha \sqsubseteq \alpha'$ and $\mathrm{dom}(\alpha') \smallsetminus \mathrm{dom}(\alpha)$ is a singleton set.  In this case, we write $\alpha'=\alpha+(\overrightarrow{m}\mapsto t)$, where $\{\overrightarrow{m}\}=\mathrm{dom}(\alpha') \smallsetminus \mathrm{dom}(\alpha)$ and $t=\alpha'(\overrightarrow{m})$.
\item The notation $\alpha \underset{\tau,T}{\overset{1}{\longrightarrow}} \alpha'$ means that $\alpha,\alpha' \in \mathcal{A}^{\tau}_T$ and $\alpha'$ is a single-tile extension of $\alpha$.  (The ``1'' above the arrow is to denote that a single tile is added at this step.)
\end{enumerate}
\end{definition}
\begin{definition}
Let $\alpha \in \mathcal{A}^{\tau}_T$.
\begin{enumerate}
\item For each $t \in T$, the \emph{$\tau$-$t$-frontier} of $\alpha$ is the set
\begin{displaymath}
\partial^{\tau}_T \alpha = \Big\{ \overrightarrow{m} \in \mathbb{Z}^2 \smallsetminus \mathrm{dom}(\alpha) \Big| \sum_{\overrightarrow{u} \in U_2} \mathrm{str}_t(\overrightarrow{u}) \cdot \llbracket \alpha(\overrightarrow{m} + \overrightarrow{u})(-\overrightarrow{u})=t(\overrightarrow{u}) \rrbracket \geq \tau \Big\} \enspace.
\end{displaymath}
\item The \emph{$\tau$-frontier} of $\alpha$ is the set
\begin{displaymath}
\partial^{\tau} \alpha = \bigcup_{t \in T} \partial^{\tau}_t \alpha \enspace.
\end{displaymath}
\end{enumerate}
\end{definition}
\begin{definition}
A \emph{$\tau$-$T$-assembly sequence} is a sequence $\overrightarrow{\alpha}=(\alpha_i \mid 0 \leq i <k)$ in $\mathcal{A}^{\tau}_T$, where $k \in \mathbb{Z}^+ \cup \{\infty\}$ and, for each $i$ with $1 \leq i+1<k$, $\alpha_i \underset{\tau,T}{\overset{1}{\longrightarrow}} \alpha_{i+1}$.
\end{definition}
\begin{definition}
The \emph{result} of a $\tau$-$T$-assembly sequence $\overrightarrow{\alpha}=(\alpha_i \mid 0 \leq i <k)$ is the unique $T$-configuration $\alpha=\mathrm{res}(\overrightarrow{\alpha})$ satisfying: $\mathrm{dom}(\alpha)=\cup_{0 \leq i<k} \mathrm{dom}(\alpha_i)$ and $\alpha_i \sqsubseteq \alpha$ for each $0 \leq i<k$.
\end{definition}
\begin{definition}
Let $\alpha,\alpha' \in \mathcal{A}^{\tau}_T$.  A \emph{$\tau$-$T$-assembly sequence from $\alpha$ to $\alpha'$} is a $\tau$-$T$-assembly sequence $\overrightarrow{\alpha}=(\alpha_i \mid 0 \leq i <k)$ such that $\alpha_0=\alpha$ and $\mathrm{res}(\overrightarrow{\alpha})=\alpha'$.  We write $\alpha \underset{\tau,T}{\longrightarrow} \alpha'$ to indicate that there exists a $\tau$-$T$-assembly from $\alpha$ to $\alpha'$.
\end{definition}
\begin{definition}
An assembly $\alpha \in \mathcal{A}^{\tau}_T$ is \emph{terminal} if $\partial^{\tau} \alpha = \emptyset$.
\end{definition}
Intuitively, a configuration is a set of tiles that have been placed in the plane, and the configuration is stable if the binding strength at every possible cut is at least as high as the temperature of the system.  Informally, an assembly sequence is a sequence of single-tile additions to the frontier of the assembly constructed at the previous stage.  Assembly sequences can be finite or infinite in length.  We are now ready to present a definition of a tile assembly system.
\begin{definition}
Write $\mathcal{A}^{\tau}_T$ for the set of configurations, stable at temperature $\tau$, of tiles whose tile types are in $T$.  A \emph{tile assembly system} is an ordered triple $\mathcal{T}=(T,\sigma,\tau)$ where $T$ is a finite set of tile types, $\sigma \in \mathcal{A}^{\tau}_T$ is the \emph{seed assembly}, and $\tau \in \mathbb{N}$ is the \emph{temperature}.  We require $\mathrm{dom}(\sigma)$ to be finite.
\end{definition}
\begin{definition}
Let $\mathcal{T}=(T,\sigma,\tau)$ be a tile assembly system.
\begin{enumerate}
\item Then the \emph{set of assemblies produced by $\mathcal{T}$} is
\begin{displaymath}
\mathcal{A}[\mathcal{T}] = \big\{\alpha \in \mathcal{A}^{\tau}_T \big| \sigma \underset{\tau,T}{\longrightarrow} \alpha \big\} \enspace,
\end{displaymath}
where ``$\sigma \underset{\tau,T}{\longrightarrow} \alpha$'' means that tile configuration $\alpha$ can be obtained from seed assembly $\sigma$ by a legal addition of tiles.
\item The \emph{set of terminal assemblies produced by $\mathcal{T}$} is
\begin{displaymath}
\mathcal{A}_{\Box}[\mathcal{T}] = \{\alpha \in \mathcal{A}[\mathcal{T}] \mid \alpha \textrm{ is terminal}\} \enspace,
\end{displaymath}
where ``terminal'' describes a configuration to which no tiles can be legally added.
\end{enumerate}
\end{definition}
If we view tile assembly as the programming of matter, the following analogy is useful: the seed assembly is the input to the computation; the addition of tile types to the growing assembly are the legal steps the computation can take; the temperature is the primary inference rule of the system; and the terminal assemblies are the possible outputs.

We are, of course, interested in being able to \emph{prove} that a certain tile assembly system always achieves a certain output.  In~\cite{solwin}, Soloveichik and Winfree presented a strong technique for this: local determinism.

Informally, an assembly sequence $\overrightarrow{\alpha}$ is locally deterministic if (1) each tile added in $\overrightarrow{\alpha}$ binds with the minimum strength required for binding; (2) if there is a tile of type $t_0$ at location $\overrightarrow{m}$ in the result of $\alpha$, and $t_0$ and the immediate ``OUT-neighbors'' of $t_0$ are deleted from the result of $\alpha$, then no other tile type in $\mathcal{T}$ can legally bind at $\overrightarrow{m}$; the result of $\alpha$ is terminal.  We formalize these points as follows.
\begin{definition}[Soloveichik and Winfree~\cite{solwin}]
A $\tau$-$\mathcal{T}$-assembly sequence $\overrightarrow{\alpha}=(\alpha_i \mid 0 \leq i \leq k)$ with result $\alpha$ is \emph{locally deterministic} if it has the following three properties.
\begin{enumerate}
\item For all $\overrightarrow{m} \in \textrm{dom}(\alpha) - \textrm{dom}(\alpha_0$),
\begin{displaymath}
\sum_{\overrightarrow{u} \in \mathrm{IN}^{\overrightarrow{\alpha}}(\overrightarrow{m})} \mathrm{str}_{\alpha_{i_{\alpha}(\overrightarrow{m})}}(\overrightarrow{m},\overrightarrow{u}) = \tau \enspace,
\end{displaymath}
where $\mathrm{IN}^{\overrightarrow{\alpha}}(\overrightarrow{m})$ means the sides of the tile that bound at location $\overrightarrow{m}$ during assembly sequence $\overrightarrow{\alpha}$ that contributed nonzero strength during the stage at which the tile bound.  (Informally, these are the ``input sides'' of the tile at location $\overrightarrow{m}$, with respect to assembly sequence $\overrightarrow{\alpha}$.)
\item For all $\overrightarrow{m} \in \textrm{dom}(\alpha) - \textrm{dom}(\alpha_0)$ and all $t \in \mathcal{T} - \{\alpha(\overrightarrow{m})\}$, $\overrightarrow{m} \notin \partial_t^{\tau}(\overrightarrow{\alpha}\backslash \overrightarrow{m})$.
\item $\partial^{\tau}\alpha =\emptyset$.
\end{enumerate}
\end{definition}
\begin{definition}[Soloveichik and Winfree~\cite{solwin}]
A tile assembly system $\mathcal{T}$ is \emph{locally deterministic} if there exists a locally deterministic $\tau$-$\mathcal{T}$-assembly sequence $\alpha=(\alpha_i \mid 0 \leq i < k)$ with $\alpha_0=\sigma$.
\end{definition}

Local determinism is important because of the following result.
\begin{thm}[Soloveichik and Winfree~\cite{solwin}]
If $\mathcal{T}$ is locally deterministic, then $\mathcal{T}$ has a unique terminal assembly.
\end{thm}
\subsection{More general self-assembly models}
We move now from DNA tiles self-assembling on a two-dimensional surface, to a more general setting, where self-assembling ``agents'' with the ability not just to bind but also to communicate after binding and potentially unbind, can assemble either in the plane or in three-space.  One could think of think of these agents as (nano- or macroscale) robots that interlock physically, and, after interlocking, can send their neighbors electronic messages of low complexity.  Based on receipt of messages, the robots can then decide to break bonds to one or more of their neighbors.  Such modular robots have already been implemented in laboratory experiments~\cite{klavins:psa}.  Further, these robots may be constructed so each has (at least with high probability) a unique identification code---permitting transmission of strictly more information than is possible in the setting of tile self-assembly, in which tiles do not have unique identifiers.

We will consider generalizations of the abstract Tile Assembly Model that include the following: (1) multiple nucleation; (2) assembly in which glues bind incorrectly according to some error probability; and (3) negative glue strengths, allowing incorrectly bound tiles to be released from the assembly so it is possible for a correctly-binding tile to attach in that location; (4) a third spatial dimension; and (5) tiles can now be ``agents,'' \emph{i.e.}, finite state machines with algorithms and unique identifiers.  We formalize this as follows.%
\begin{definition}
A \emph{$d$-regular self-assembling agent type} $T$ is a finite state machine of form $T=\langle A,(g_1,\ldots,g_d) \rangle$, where $A$ is an (deterministic or probabilistic) algorithm and the $g_i$'s are finite strings over a finite alphabet (codes for the glue types associated with $T$) that are hardcoded into the machine.  The algorithm $A$ can be null (in the case of passive self-assembly like DNA tiles), or can decide whether to transmit messages of length bounded by a constant to neighbors based on the agent's interaction with neighboring glue types.
\end{definition}
We will assume that all agent types have identical geometric structure, and their $d$ glues all have the same orientation. For example, in the aTAM, all agent types are unit squares, oriented north, east, south, west.  Also, for simplicity of the proof, we will assume that our agents are memoryless.  However, because of the generality of the results of Naor and Stockmeyer, our lower bound results would still hold if agents could make active self-assembly decisions based on a history of messages received from neighbors, not just the current messages and glue types of their neighbors.
\begin{definition}
The binary relation $R$ is a \emph{set of binding rules} for the (finite) set of agent types $\{T_i\}_i$ if, for any $(x,y) \in R$, both $x$ and $y$ are glue types that appear in elements of $\{T_i\}_i$.
\end{definition}
\begin{definition}
The function $\beta$ is \emph{an assignment of binding strengths} for the set of binding rules $R$, if the domain of $\beta$ is $R$, and the range of $\beta$ is the set of nonnegative integers.
\end{definition}
\begin{definition}
$\mathcal{M}$ is a \emph{model of $d$-regular self-assembling agents} if $\mathcal{M}=\langle \{T_i\}_i,R,\beta,\tau, \sigma \rangle$, where $\{T_i\}_i$ is a (finite) set of $d$-regular self-assembling agent types, $R$ is a set of binding rules for $\{T_i\}_i$, $\beta$ is an assignment of binding strengths for $R$, $\tau$ is the temperature of the system (the threshold binding strength for bonds to be stable), and $\sigma$ is an initial (finite) seed assembly.
\end{definition}
The algorithm of each agent type may include a variable MY-ID, and we allow the possibility that each agent in the system does, in fact, have a unique identification number.  This might be appropriate when modeling robotic self-assembly.  In the case of molecular self-assembly, each agent is anonymous.  Assembly systems in both the aTAM and the stochastic model of Majumder \emph{et al.} can be defined in this formalism, by giving each agent type an algorithm that performs no instructions, and defining (respectively deterministic or probabilistic) binding relations in a natural way.

To conclude this section, we formalize what it means for a self-assembly model to allow multiple nucleation on a surface.
\begin{definition}
Let $\mathcal{M}$ be a model of $d$-regular self-assembling agents.  We say $\mathcal{M}$ \emph{allows multiple nucleation} if, in addition to the placement of the seed assembly at the initial stage of assembly, there is some probability $\pi_{\nu}$ that (at the first stage of assembly only) an agent is placed on each location of the surface with probability $\pi_{\nu}$.  Further, if an agent is placed at location $\overrightarrow{m}$ because of multiple nucleation, its agent type is chosen uniformly at random from the space of possible agent types.
\end{definition}
We could allow multiple nucleation to occur at multiple stages during the assembly, not just the first.  Again, because of the generality of Naor and Stockmeyer's results, that would not affect our lower bound proof.
\section{Distributed Computing Results of Naor and Stockmeyer}
In a well known distributed computing paper, Naor and Stockmeyer investigated whether ``locally checkable labeling'' problems could be solved over a network of processors in an entirely local manner, where a local solution means a solution arrived at ``within time (or distance) independent of the size of the network''~\cite{ns}.  One locally checkable labeling problem Naor and Stockmeyer considered was the \emph{weak $c$-coloring problem}.

\begin{definition}[Naor and Stockmeyer~\cite{ns}]
For $c \in \mathbb{N}$, a \emph{weak $c$-coloring} of a graph is an assignment of numbers from $\{1,\ldots,c\}$ (the possible ``colors'') to the vertices of the graph such that for every non-isolated vertex $v$ there is at least one neighbor $w$ such that $v$ and $w$ receive different colors.  Given a graph $G$, the \emph{weak $c$-coloring problem for $G$} is to weak $c$-color the nodes of $G$.
\end{definition}

In the context of tiling, to solve the weak $c$-coloring problem for an $n \times n$ surface means tiling the surface so each tile has at least one neighbor (to the north, south, east or west) of a different color.  In the next section, we will present a simple solution to the weak $c$-coloring problem in the abstract Tile Assembly Model.  By contrast, Naor and Stockmeyer showed that no local, constant-time algorithm can solve the weak $c$-coloring problem for grid graphs, nor for $k$-dimensional meshes, a generalization of grid graphs which we now define.
\begin{definition}
A \emph{$k$-dimensional mesh} is a graph with vertex set $\{0,1,\ldots,m\}^k$ for some $m$, such that two vertices are connected by an edge if the $\mathcal{L}_1$-distance between them is 1.
\end{definition}
\begin{thm}[Naor and Stockmeyer~\cite{ns}] \label{theorem:ns1}
For any natural numbers $c$, $k$ and $t$, there is no local algorithm with time bound $t$ that solves the weak $c$-coloring problem for the class of $k$-dimensional meshes.  (This remains true even if the processors have unique identifiers and can transmit them as part of the local algorithm.)
\end{thm}
%\begin{theorem}[Naor and Stockmeyer~\cite{ns}] \label{theorem:ns1}
%For any $c$ and $t$, there is no local algorithm with time bound $t$ that solves the weak $c$-coloring problem for the class of finite square grid graphs over the integer lattice.
%\end{theorem}
%This theorem is a consequence of Theorem 6.3 in~\cite{ns}.  The original result is a stronger statement.
A second theorem from the same paper says that randomization does not help.  The original result is stronger than the formulation here.
\begin{thm}[Naor and Stockmeyer~\cite{ns}] \label{theorem:ns2}
Fix a class $\mathcal{G}$ of graphs closed under disjoint union.  If there is a randomized local algorithm $P$ with time bound $t$ that solves the weak $c$-coloring problem for $\mathcal{G}$ with error probability $\epsilon$ for some $\epsilon<1$, then there is a deterministic local algorithm $A$ with time bound $t$ that solves the weak $c$-coloring problem for $\mathcal{G}$.
\end{thm}

\section{Simulation of Self-Assembly on a Surface}
In order to apply the theorems of Naor and Stockmeyer to the realm of self-assembly, we build a distributed network of processors that reduces a self-assembly problem to a distributed computing problem.  The motivating intuition is that each processor simulates a location of the surface, and reports to its neighbors whether there is a (simulated) agent at that location.  Formally, we prove the following theorem.
\begin{thm} \label{theorem:simulation}
Let $\mathcal{M}$ be a model of $d$-regular self-assembling agents for any natural number $d>0$, such that $\mathcal{M}$ self-assembles on a $k$-dimensional mesh of size $n^k$, and that $\mathcal{M}$ allows multiple nucleation.  Then there is a model $\mathcal{N}$ of distributed computing that simulates $\mathcal{M}$ using $n^k$ processors with the network topology of a $k$-dimensional mesh, and constant-size message complexity.
\end{thm}
\begin{proof}
Fix a model of $d$-regular self-assembling agents $\mathcal{M}$ as in the theorem statement.  Let $\alpha$ be a configuration of agents on the mesh of size $n^k$.  Let $\Gamma$ be the set of glue types of $\mathcal{M}$ and $M$ the set of electronic messages of $\mathcal{M}$.  (Both $\Gamma$ and $M$ are finite sets.)  The definition of the binding function $\beta$ induces a function
\begin{displaymath}
\hat{\beta}: (T \cup \{ \emptyset \} ) \times ( \Gamma \cup \{ \emptyset \} )^d \times (M \cup \{ \emptyset \} )^d \times (T \cup \{ \emptyset \} ) \longrightarrow [0,1]
\end{displaymath}
such that $\hat{\beta}$ takes as input the (possibly empty) agent type $\alpha(\overrightarrow{m})$ at some location $\overrightarrow{m}$ in configuration $\alpha$, and, based on the glue types and electronic messages received from the $d$ neighbors that could be incident to an agent at $\overrightarrow{m}$, returns, for each agent type, the probability that $\alpha(\overrightarrow{m})$ would contain that agent type, over the space of all legal $\mathcal{M}$-assembly sequences that start with configuration $\alpha$ and run for one time step.  In particular, for fixed $t_0 \in T$, fixed $\gamma_1,\ldots,\gamma_d \in \Gamma \cup \{ \emptyset \}$, and fixed $m_1,\ldots,m_d \in M \cup \{ \emptyset \}$, it is true that
\begin{displaymath}
\sum_{t \in T \cup \{ \emptyset \} } \hat{\beta}(t_0,\gamma_1,\ldots,\gamma_d,m_1,\ldots,m_d,t) = 1 \enspace .
\end{displaymath}
We have not formally defined $\mathcal{M}$-assembly sequences, but they are a natural extension of the $\tau$-$T$-assembly sequences of tile self-assembly, where the $\beta$ and $\tau$ of $\mathcal{M}$ are used to determine whether agents bind stably to one another.  Also, if an agent type lies on the edge of the $n^k$-size surface, so it does not have a neighbor in a particular direction, we define $\hat{\beta}$ so that the empty set is the glue and electronic message ``transmitted'' from the ``neighbor'' in that direction.

We simulate assembly sequences of $\mathcal{M}$ on an $k$-dimensional mesh where each of the dimensions has length $n$ by a network of processors $\mathcal{N}$ whose network graph is also a $k$-dimensional mesh of total size $n^k$.  Each processor will simulate the presence or absence of an agent in the same location on the assembly surface.  We interpret bonds between two agents as messages.  We add on top of those messages, an additional set of electronic messages agents can send neighbors, and encode the combination as an ordered pair: glue type and electronic message.  The function $\hat{\beta}$ will be the probabilistic transition function for processors in this system.

Processors of $\mathcal{N}$ are of the following form.

\begin{description}
\item[\underline{Processor $p_i$}]
\item[$d$-many input message buffers:] $\mathrm{inbuf}_{i,1}, \ldots, \mathrm{inbuf}_{i,d}$.
\item[$d$-many output message buffers:] $\mathrm{outbuf}_{i,1}, \ldots, \mathrm{outbuf}_{i,d}$.
\item[A color variable:] $\mathrm{COLOR}_i$, a variable that can take a value from $\{1,\ldots,c\}$, where $c$ is a global constant.
\item[A local state:] Each processor is in one of $|T|+1$ different local states $q$ during a given execution stage $s$.  There is one stage $q_k$ to simulate each agent type $t_k \in T$, and an additional stage EMPTY, to simulate the absence of an agent from the surface location that $p_i$ is simulating.
\item[A state transition function:] This function takes the current processor state and the messages received in the current round, and probabilistically directs what state the processor will adopt in the next round.
\end{description}

The messages processors send on the network are of form $\langle$glue type, electronic message$\rangle$.  The input message buffers of processor $p_i$ simulate the glue types of the edges the agent at $p_i$'s location is adjacent to, and the electronic messages (if any) received from an agent's neighbors.  The output message buffers of $p_i$ simulate the glues on the edges of the tile $p_i$ is simulating, and the electronic messages the agent transmits to its neighbors.  The purpose of $\mathrm{COLOR}_i$ is to simulate the color of the agent placed at the location simulated by $p_i$.

All processors in $\mathcal{N}$ are hardcoded with the same probabilistic state transition function, which is determined from the definition of $\hat{\beta}$ (which we induced above from the properties of $\mathcal{M}$), in the natural way: if, in round $r$ of the algorithm execution, $p_i$ is in state $q_k$, a simulation of $t_k \in T$, and hears messages that simulate glue types $g_1,\ldots,g_d$ and electronic messages $m_1,\ldots,m_d$, then at the end of round $r$, it will transition to state $q_j$ with probability $\pi_j$, where $\hat{\beta}(t_k,g_1,\ldots,g_d,m_1,\ldots,m_d,t_j)=\pi_j$ and each $t_j$ is a distinct element of $T \cup \{ \emptyset \}$.  As explained above, we denote the state that simulates the ``presence of the empty set''---\emph{i.e.,} the absence of any agent from the location simulated by $p_i$---as EMPTY.

To simulate the process of self-assembly, we run the following distributed algorithm on $\mathcal{N}$.

Algorithm execution proceeds in synchronized rounds.  Before execution begins, all processors start in state EMPTY.  In round $r=0$, (through the intervention of an omniscient operator) each processor in the locations corresponding to the seed assembly enters the stage to simulate the agent type at that location in the seed assembly.

Also in round $r=0$, each processor not simulating part of the seed assembly ``wakes up'' (enters a state other than EMPTY) with probability $\pi_{\nu}$, the multiple nucleation probability of $\mathcal{M}$.  If a processor wakes up, it enters state $q \neq \mathrm{EMPTY}$, chosen uniformly at random from the set of non-EMPTY states.  For any round $r>0$, each processor runs either Algorithm 1 or Algorithm 2, depending on whether it is in state EMPTY.
%\begin{description}
%\item[Stage 1] Each processor in state QUIET, with probability $\pi_1$ will enter a state that simulates the presence of an agent type appearing (``nucleating'') at that location in structure $S$.  Simultaneous with this, any processor in state QUIET with neighbors such that one or more agent types could stably bind to the corresponding location of $S$, will change state with probability $\pi_2$, to a state that simulates the presence of a stably bound agent in that location.  (The values of $\pi_1$ and $\pi_2$ depend on the behavior of the self-assembly we wish to simulate.)
%\item[Stage 2] Each processor not in state QUIET empties its input buffer, and runs the algorithm defined by the agent type its state is simulating.  If the algorithm directs that the agent severs contact with one or more neighbors, the processor sends a message to that set of neighbors saying that its binding strength to those neighbors is now zero.
%\item[Stage 3] Any processors simulating agents not stably bound at this point in the round change their states to QUIET.
%\end{description}
\begin{algorithm}
\caption{For $p_i$ in state EMPTY at round $r$}
\begin{algorithmic}
\IF{$r=0$}
\STATE{wake up with probability $\pi_{\nu}$, and cease execution for this round.}
\ENDIF
\IF{$r>0$}
\STATE{Read the $d$-many input buffers.}
\IF{no messages were received}
\STATE{cease execution for the round}
\ELSE
\STATE{let $q_0$ be the state change obtained according to probabilities $\hat{\beta}$ assigns to the space $T \cup \{ \emptyset \}$, for a location that has adjacent glue types and electronic messages that are simulated by the messages received this round.}
\STATE{Send the messages indicated by state $q_0$ and the behavior of $A$.}
\STATE{Set the value of $\mathrm{COLOR}_i$ according to $q_0$.}
\STATE{Enter state $q_0$ and cease execution for this round.}
\ENDIF
\ENDIF
\end{algorithmic}
\end{algorithm}
\begin{algorithm}
\caption{For $p_i$ in state $q \neq \mathrm{EMPTY}$ (at any round)}
\begin{algorithmic}
\STATE{Read the four input buffers.}
\IF{no messages were received}
\STATE{Send the messages indicated by state $q$ and the behavior of $A$ and cease execution for this round.}
\ELSE
\STATE{Let $q_0$ be the state change obtained probabilistically, based on the probabilities produced by the function $\hat{\beta}$ to the space $T \cup \{ \emptyset \}$, given input from the glue types and electronic messages simulated by the messages received this round.}
\STATE{Send the messages indicated by state $q_0$.}
\STATE{Set the value of $\mathrm{COLOR}_i$ according to $q_0$.}
\STATE{Enter state $q_0$ and cease execution for this round.}
\ENDIF
\end{algorithmic}
\end{algorithm}

The interaction between agents in $\mathcal{M}$ is completely defined by the glues and electronic messages of an agent's immediate neighbors, as specified in the function $\hat{\beta}$ and the algorithm of each agent type.  The processors of $\mathcal{N}$ simulate that behavior with Algorithm 2.  Since the processors of $\mathcal{N}$ simulate empty locations with Algorithm 1, by a straightforward induction argument, $\mathcal{N}$ can simulate all possible $\mathcal{M}$-assembly sequences, and the theorem is proved.
\end{proof}

We obtain our time lower bound results as corollaries of Theorem~\ref{theorem:simulation}.
\begin{figure}
\centering
\includegraphics[height=3in]{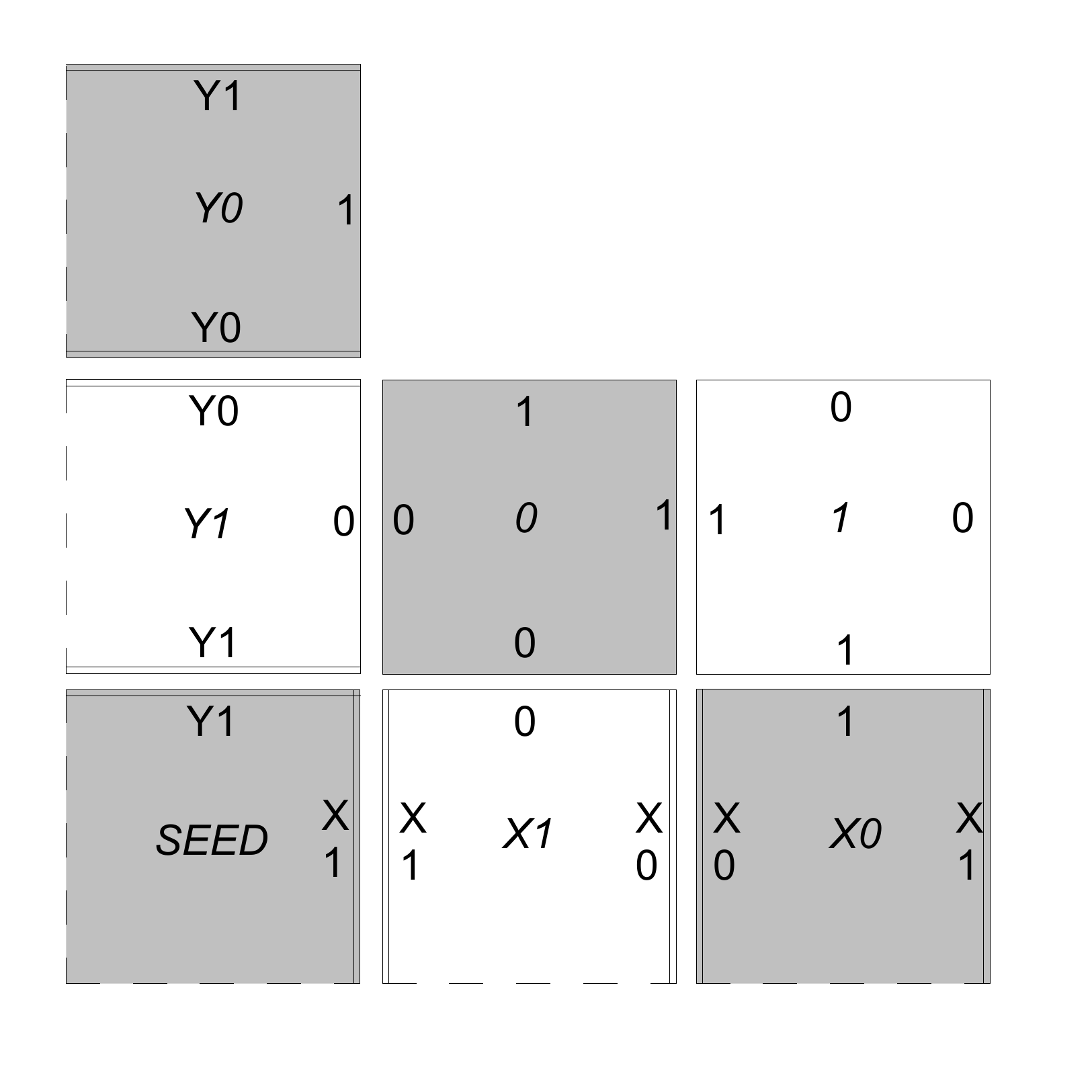}
\caption{The tileset $\mathcal{T^*}$ used in the proof of Proposition 1.} \label{figure:tileset}
\end{figure}

\begin{cor}
If the (deterministic or probabilistic) binding rules of a multiply nucleating tile assembly system $\mathcal{T}$ are entirely local, then $\mathcal{T}$ is unable to solve the weak $c$-coloring problem in constant time.
\end{cor}
\begin{proof}
Suppose $\mathcal{T}$ is an irreversible tiling model.  If $\mathcal{T}$ can weak $c$-color surfaces in constant time, then there is a deterministic algorithm for the distributed network $\mathcal{N}$ that weak $c$-colors $\mathcal{N}$ locally, and in constant time.  By Theorem~\ref{theorem:ns1} that is impossible.

So assume $\mathcal{T}$ is a reversible tiling model, and when $\mathcal{T}$ assembles, it weak $c$-colors the tiling surface, and achieves bond pair equilibrium in constant time.  Then there is a local probabilistic algorithm for $\mathcal{N}$ that weak $c$-colors $\mathcal{N}$ in constant time, with positive probability of success.  By Theorem~\ref{theorem:ns2} that is impossible as well.  Therefore, no $\mathcal{T}$ exists that weak $c$-colors surfaces in constant time.
\end{proof}
By a similar argument, we obtain a lower bound for active self-assembling agents on a three-dimensional cubic grid.
\begin{cor} \label{corollary:3D}
If a model of 6-regular self-assembling agents has only local binding rules, then it cannot solve the weak $c$-coloring problem in constant time on a 3-dimensional mesh, for any value of $c$.
\end{cor}
A physical interpretation of Corollary~\ref{corollary:3D} would be that robots self-assembling in three-space (\emph{i.e.}, $k=3$ and, in this example, $d=6$ so there are six arms coming off each robot, orthogonally to one another) cannot achieve speedup to constant time by self-assembling in separate groups and then joining the groups together.  This lower bound remains in effect even if the robots are designed by a method that assigns each robot a unique identifier.

We conclude this section by noting that the weak $c$-coloring problem has low tile complexity---that is, can be defined using only a few local rules---in the aTAM.
\begin{prop}
There is a tile assembly system in the abstract Tile Assembly Model that weak $c$-colors the first quadrant, using only seven distinct tile types.
\end{prop}
\begin{proof}
Figure 2 exhibits a tileset $\mathcal{T}^*$ of seven tile types that assembles into a weak $c$-coloring of the first quadrant, starting from an individual seed tile placed at the origin.  One can verify by inspection that $\mathcal{T}^*$ is locally deterministic, so it will always produce the same terminal assembly.  All assembly sequences generated by $\mathcal{T}^*$ produce a checkerboard pattern in which a monochromatic ``+'' configuration never appears.  Hence, it solves the weak $c$-coloring problem for the entire first quadrant, and also for all $n \times n$ squares, for any $n$.
\end{proof}
One can define a three-dimensional version of the tileset $\mathcal{T}^*$ (shown in Figure~\ref{figure:tileset}) in the natural way, using for example the 3D tile assembly model in~\cite{3D}.  Such a three-dimensional tileset will weakly $c$-color the three-dimensional mesh where $d=6$, with low tile complexity.
\section{Conclusion}
In this paper, we showed that if a tile assembly model has only local binding rules, then it cannot use multiple nucleation on a surface to solve locally checkable labeling problems in constant time, even though the abstract Tile Assembly Model can solve a locally checkable labeling problem using just seven tile types.  In fact, we proved a more general impossibility result, which showed the same lower bound applies to self-assembling agents in a three-dimensional grid that are capable of binding and subsequently sending messages to their neighbors.  To the best of our knowledge, this was the first application of a distributed computing impossibility result to the field of self-assembly.

There are still many open questions regarding multiple nucleation.  Aggarwal \emph{et al.} asked in~\cite{complexities} whether multiple nucleation might reduce the tile complexity of finite shapes.  The answer is not known.  Furthermore, we can ask for what class of computational problems does there exist some function $f$ such that we could tile an $n \times n$ square in time $\mathcal{O}(1)<\mathcal{O}(f)<\mathcal{O}(n^2)$, and ``solve'' the problem with ``acceptable'' probability of error, in a tile assembly model that permits multiple nucleation.  It would also be interesting to explore the possibility of modeling multiple nucleation of molecules floating in solution---instead of adhering to a surface---perhaps by using techniques from the field of \emph{ad hoc} wireless networks.

We hope that this is just the start of a conversation between researchers in the fields of distributed computing and biomolecular computation.

\section*{Acknowledgements}
I am grateful to Soma Chaudhuri, Dave Doty, Jim Lathrop and Jack Lutz for helpful discussions on earlier versions of this paper.  I am also grateful to two anonymous referees, who suggested significant conceptual and technical improvements to my original journal submission.

\end{document}